\documentclass{article}

\usepackage{PRIMEarxiv}

\usepackage[utf8]{inputenc} % allow utf-8 input
\usepackage[T1]{fontenc}    % use 8-bit T1 fonts
\usepackage{hyperref}       % hyperlinks
\usepackage{url}            % simple URL typesetting
\usepackage{booktabs}       % professional-quality tables
\usepackage{amsfonts}       % blackboard math symbols
\usepackage{nicefrac}       % compact symbols for 1/2, etc.
\usepackage{microtype}      % microtypography
\usepackage{lipsum}
\usepackage{fancyhdr}       % header
\usepackage{graphicx}       % graphics
\graphicspath{{media/}}     % organize your images and other figures under media/ folder
\usepackage{hanging}  
\usepackage{multirow}
\usepackage{longtable}
\usepackage{array}
\usepackage{float}

\title{Evaluating Investment Risks in LATAM AI Startups: Ranking of Investment Potential and Framework for Valuation

\thanks{\textit{\underline{Citation}}: 
\textbf{Ramos, Abraham and Montoya, Laura N. Evaluating Investment Risks in LATAM AI Startups: Ranking of Investment Potential and Framework for Valuation.}} 
}

\author{
  Abraham Ramos-Torres \\
  Accel AI Institute \\
  Universidad Central de Venezuela \\
  Caracas, Venezuela\\
  \texttt{abraham@accel.ai} \\
   \And
  Laura N. Montoya \\
  Accel AI Institute \\
  San Francisco, CA\\
 \texttt{laura@accel.ai} \\
}

\begin{document}
\maketitle

\begin{abstract}
The growth of the tech startup ecosystem in Latin America (LATAM) is driven by innovative entrepreneurs addressing market needs across various sectors. However, these startups encounter unique challenges and risks that require specific management approaches. This paper explores a case study with the Total Addressable Market (TAM), Serviceable Available Market (SAM), and Serviceable Obtainable Market (SOM) metrics within the context of the online food delivery industry in LATAM, serving as a model for valuing startups using the Discounted Cash Flow (DCF) method. By analyzing key emerging powers such as Argentina, Colombia, Uruguay, Costa Rica, Panama, and Ecuador, the study highlights the potential and profitability of AI-driven startups in the region through the development of a ranking of emerging powers in Latin America for tech startup investment. The paper also examines the political, economic, and competitive risks faced by startups and offers strategic insights on mitigating these risks to maximize investment returns. Furthermore, the research underscores the value of diversifying investment portfolios with startups in emerging markets, emphasizing the opportunities for substantial growth and returns despite inherent risks.
\end{abstract}

% keywords can be removed

\section{Introduction}
Valuation of startups is a key component of investment decisions, it becomes more important when it comes to emerging countries, particularly in the venture capital and private equity sectors. Traditional valuation methods often fall short when applied to tech startups due to their unique characteristics, such as high uncertainty and rapid growth potential. This is because tech startups operate in dynamic and rapidly changing markets, making it difficult to accurately predict their future performance and value using traditional valuation models. The evaluation of young companies proves to be a complex and delicate task which often includes a high level of uncertainty, lack of history, low survival rates, and lack of balance sheet figures, which generate special valuation problems \cite{Derin2018EvaluationMethod}.

In Latin America, the tech industry has seen substantial growth in recent years, largely due to increasing entrepreneurial activities. Countries like Panama, Chile, Guatemala, and Brazil have been particularly active in early-stage entrepreneurial endeavors, contributing to the overall growth of the tech industry. Additionally, investment in artificial intelligence (AI) is particularly promising, with the potential to significantly impact the region's real GDP. This demonstrates the region's growing focus on innovation and technology, which is crucial for driving economic development and creating new opportunities for businesses and individuals.

However, investing in tech startups in Latin America carries significant risks due to the region's instability, encompassing economic, financial, and political factors. To mitigate these risks, it is crucial to adopt a comprehensive approach that assesses each of the inherent risks of startups and develops specific risk management strategies.

Emerging markets present unique opportunities for high returns due to rapid economic growth and evolving landscapes \cite{Kohers1998TheStrategies}. Adding emerging markets to portfolios can enhance risk-return profiles and provide similar diversification benefits to more complex strategies \cite{Kohers1998TheStrategies}\cite{Beach2006WhyPortfolio}. Some studies indicate that portfolios focusing solely on emerging markets can perform as well as those that combine emerging and developed markets \cite{Arora2011InternationalStrategy}. Specifically, Latin American markets in Colombia, Brazil, Chile, and Mexico show a positive link between risk and return in Latin America, although this is not always perfectly true \cite{Tsuji2013AnMarkets}.  Overall, emerging market equities, including those in Latin America, can improve portfolio performance by offering superior returns and enhancing risk-adjusted measures such as the Sharpe ratio and reward-to-semideviation ratio \cite{Beach2006WhyPortfolio}.

This study aims to provide a framework for converting investability into valuation for tech startups in Latin America. Through a detailed analysis of the infrastructure, education, and economic factors necessary to launch and scale AI companies in the region, this study offers a ranking of the most promising Latin American countries for AI investment. Using metrics such as TAM (Total Addressable Market), SAM (Serviceable Available Market), and SOM (Serviceable Obtainable Market), the growth potential and market profitability for startups in the online food industry applying artificial intelligence are estimated.

The study emphasizes the potential profitability and investment attractiveness of startups in the online food industry in countries such as Mexico, Colombia, and Brazil. It also examines the specific challenges faced by countries like Guatemala, Nicaragua, Venezuela, Cuba, and Honduras, due to their high levels of poverty, political instability, and economic challenges. These factors limit the profitability of startups in these markets.

\section{INVESTABILITY FACTORS IN LATIN AMERICA}

The technology industry in Latin America has seen substantial growth in recent years. In 2022, the total value of the tech industry in Latin America reached US\$517 billion, with a growth rate of 5.6\% \cite{Grootjen2023An2023}. Furthermore, this growth is closely linked to entrepreneurial activities in the region. Latin American countries are among the countries with the highest entrepreneurship rates in the world. Leaders include Panama 31\%, Chile 31\%, Guatemala 32\%, and Brazil 19\% of adults aged 18 to 64 engaged in early-stage entrepreneurial activity (TEA) in 2023 (Global Entrepreneurship Monitor, 2023).

The potential for growth in the technology sector in Latin America is immense, especially for startup investments. In the field of artificial intelligence (AI), it is expected to have a positive impact on the real GDP of the region by 5.4\% by 2030, which amounts to approximately US\$0.5 trillion \cite{Dr.AnandS.RaoSizingCapitalise}. However, it is important to be cautious when dealing with risky investments. Before making an investment decision, several factors need to be considered. This ranking was created using a previous version of the study's ranking as a foundation and then updated with more recent and additional metrics data \cite{Torres2024AIAmerica}. This new ranking of emerging powers for tech startup investments in Latin America offers a comprehensive analysis of the factors that make the region an attractive destination for investments in this technology.

The market potential for tech startups in Latin America has seen a significant increase in recent years. From 2010 to 2020, the value of the startup ecosystem has grown by 32 times, from US \$7 billion in 2010 to \$221 billion in 2020. Notably, 83\% of this growth occurred in the last four years. The startup ecosystem has created over 245,000 jobs, and more than 1,005 startups have raised US \$1 million \cite{Pena2021TecnolatinasAge}. In 2021, Latin America and the Caribbean (LAC) emerged as the region with the fastest growth in the world for venture capital (VC) investments, and in 2015, the region demonstrated bolder entrepreneurial characteristics in LATAM than in Western Europe. Only in 2024 year-to-date, VC investments into startups in countries like Colombia, Mexico, and Ecuador have reached more than US \$100 million each, and other nations such as Argentina (US \$45.3 million), Uruguay (US \$14.9 million), Peru (US \$22.8 million), and Chile (US \$21.8 million) have already illustrated that emerging powers already started receiving investments in the startup sector. This remarkable trend underscores the region's increasing attractiveness and potential for nurturing and supporting tech startups. Despite the economic challenges that have historically impacted the region, Latin American entrepreneurs have showcased remarkable adaptability, resilience, and an exceptional ability to innovate by offering new solutions to existing challenges. This has been instrumental in addressing the prevalent issues of political instability, inflation, and economic downturns that have affected the region in recent years \cite{Berg2024CatalystsAmerica}\cite{Lopes2018ComparativeAmerica}\cite{Dealroom.co2024LatinAmerica}.

The surge of VC investments in Latin America and the Caribbean not only reflects the growing confidence of investors in the region but also signifies a positive trajectory for the entrepreneurial ecosystem. As startups continue to thrive, the infusion of capital bodes well for fostering innovation, creating employment opportunities, and driving economic growth. The resiliency and ingenuity exhibited by Latin American entrepreneurs amid adversity have not only garnered attention but have also inspired confidence in the region's potential as a burgeoning hub for technological innovation. The momentum generated by the rise in VC investments is indicative of the untapped potential that exists within Latin America and the Caribbean. It serves as a testament to the robust spirit of entrepreneurship and the untiring efforts of visionary individuals who are driving positive change in the region's business landscape.

In Latin America, the talent in the AI factor has been growing in the past years, emerging powers in LATAM such as Argentina and Colombia, and current powers Brazil, Mexico, and Chile are leading the punctuation in this matter, where the AI talent concentration is higher \cite{PortulansInstitute2023Network2023}. Following this line, tech talent diversity in LATAM has been growing as well, different initiatives were developed in order to support the women's community in this sector, for example, the Tech Latinas initiative, which aims to train young women in coding and tech skills to meet the projected demand for 1.2 million web developers by 2025 \cite{Schill2021TechTechnology}. However, it is important to note that due to the economic situation that many Latin American countries are facing, there has been a loss of talent in different countries. Venezuela is one of the most affected countries, with 1783 researchers leaving the country from 1960 to 2016. At that time, that number represented 14\% of the researchers' community and was responsible for 31\% of all the research publications made in Venezuela. From another perspective, countries like Ecuador have benefited from this situation since this country was a privileged destination for immigrating researchers and engineers \cite{Requena2016PERDIDAINVESTIGADORES}.

\section{VALUATION METHODS FOR TECH STARTUPS}

Some of the most common methods to value startups include the Scorecard Method, Risk Factor Summation Method, Venture Capital Method, and Discounted Cash Flow Method \cite{Akkaya2020StartupValuation}\cite{Gonnella2023TheStartups}. However, these are not the only factors to consider when valuing a startup. Non-quantifiable variables are important to measure the potential growth of a startup, based on this, other methods such as the Berkus Methods incorporate factors not related to the financial area, for example, reputational factors, management team, business idea, and human capital \cite{Hsu2014WhatInvestors}\cite{Gonnella2023TheStartups}. Additionally, aspects such as industry, geographic location, and intellectual property need to be considered when valuing startups \cite{Berre2022WhichMicro-Targeted}. Incorporating these factors into the analysis of a startup can provide a better understanding of the environment of the organization.

\section{CONVERTING INVESTABILITY INTO VALUATION}

As mentioned above, the valuation of startups could be a challenge due to inherent complexities such as the nonexistence of historical data, their dynamic environment, and high levels of uncertainty, which are translated into risks \cite{KALINSKY2023FEATURESACTIVITY}. These unique characteristics make traditional valuation methods unsuitable for startup valuation since there must be a comprehensive understanding of business models, and also the comprehension of non-quantifiable factors that can help to measure the value across the potential showed of a startup \cite{Hsu2014WhatInvestors}\cite{Pynadath2022ValuationMethods}. For this reason, it is indispensable to include other factors in these valuation models.

Measuring the factors that make this region favorable for tech investments is not enough to understand the potential risks these types of investments can bring and how to mitigate those risks. An additional methodology is needed to manage the risks further. Applying the knowledge that the Ranking of Emerging Powers in LATAM provides, in addition to a valuation method that is focused on the market size metrics (TAM, SAM, and SOM), considering the risks associated with emerging markets, non-quantifiable and non-financial factors, and the process of converting potential and attractiveness of a startup into a quantifiable valuation, can make an investment decision more profitable. By incorporating these elements into traditional valuation methods, investors can better assess the potential and risks of the growing startup environment in Latin America.

\section{METHODOLOGY}

Our methodology for assessing the potential for AI Investment in each country included the development of a ranking that accounts for the primary indicators under the infrastructure, education, and economic elements required to launch and scale AI companies in each country successfully.

\begin{table}[ht]
  \centering
  \caption{Infrastructure elements for AI investment potential ranking}
    \begin{tabular}{p{5.39em}p{12.72em}p{9.835em}p{11.5em}}
    \toprule
    \multicolumn{4}{c}{INFRASTRUCTURE PILLAR} \\
    \midrule
    Element & Description & Indicator & Source \\
    \midrule
    Electricity & Does the country have a good implementation of electricity? & Access to Electricity (\%) & The World Bank, Access to electricity (\% of population) \cite{WorldBankOpenData2023AccessPopulation}\\
    \midrule
    \multirow{5}[30]{*}{Internet} & \multirow{5}[30]{12.72em}{Does the country have a good distribution of internet to apply AI?} & Access to Internet (\%) & World Bank Group, Internet Access and Use in Latin America and The Caribbean \cite{WorldBankOpenData2023AccessPopulation}\\
\cmidrule{3-4}    \multicolumn{1}{c}{} & \multicolumn{1}{c}{} & Internet Penetration (Urban) & World Bank Group, Internet Access and Use in Latin America and The Caribbean \cite{WorldBankGroup2022Internet2021} \\
\cmidrule{3-4}    \multicolumn{1}{c}{} & \multicolumn{1}{c}{} & Internet Penetration (Rural) & World Bank Group, Internet Access and Use in Latin America and The Caribbean \cite{WorldBankGroup2022Internet2021} \\
\cmidrule{3-4}    \multicolumn{1}{c}{} & \multicolumn{1}{c}{} & Internet Speed (Fixed Broadband) & Speedtest Global Index \cite{Speedtest2024InternetWorld}\\
\cmidrule{3-4}    \multicolumn{1}{c}{} & \multicolumn{1}{c}{} & Internet Speed (Mobile) & Speedtest Global Index \cite{Speedtest2024InternetWorld}\\
    \midrule
    Supercomputers & What is the existing supercomputer capacity within the country? & HPC Systems & RISC2 \cite{Hafner2021RISC2LATAM}\\
    \midrule
    \multirow{3}[15]{*}{Cloud Services} & \multirow{3}[15]{12.72em}{How is the existing cloud services status in LATAM?} & Number of Cloud Services & TeleGeography \cite{TeleGeography2024CloudMap}\\
\cmidrule{3-4}    \multicolumn{1}{c}{} & \multicolumn{1}{c}{} & Number of Cloud Providers & Data Center Map \cite{DataCenterMapCloudInfrastructure}\\
\cmidrule{3-4}    \multicolumn{1}{c}{} & \multicolumn{1}{c}{} & Global Cloud Ecosystem Index Score & MIT \cite{MITTechnologyReview2022Global2022}\\
    \midrule
    Models & How are the existing ML models in LATAM? & Notable ML models by geographic area (2003-2023) & Stanford University \cite{StanfordUniversity2024Artificial2024}\\
    \bottomrule
    \end{tabular}%
  \label{tab:infra-elements}%
\end{table}%

Successful implementation of artificial intelligence in Latin America requires infrastructure access including electricity, internet availability, and distribution, fast internet speed, and the availability of supercomputers in both urban and rural areas \cite{Cesareo2023TheIndex}. These factors are the most important for developing artificial intelligence (AI) solutions since the significant amount of power to operate, the essence of the data for creating unbiased models, the importance of speed when transferring large data sets, and collaboration between AI researchers and developers, and the role of the high computational power needed to implement the AI. Table I. includes the infrastructure elements and data sources included to divide the ranking.

To achieve the best possible implementation of artificial intelligence, it is crucial to approach, develop, and plan it with the help of experts in the field of computer science who can carry out large-scale projects. Therefore, education and research on artificial intelligence are considered essential factors for companies, educational institutions, and government institutions \cite{Szczepanski2019EconomicAI}. Table II includes the education and research elements and data sources used to devise the ranking.

Companies and countries must consider financial and economic factors when making decisions about investing in artificial intelligence. it's essential to ensure that the investment in AI is financially feasible and that the returns on investment are significant. Moreover, companies and countries must seek the potential to improve productivity and profitability through applying strategies based on the current economic status of the country. Risk management must be taken into account while investing in AI, and companies and countries must be aware of the potential risks involved. The possibility of innovation is another critical factor that companies and countries must consider. Investing in AI can lead to new opportunities for innovation, which can boost productivity and profitability. Understanding the competition in a specific location is essential for successful AI implementation. Companies and countries must consider the AI initiatives of their competitors and the potential impact on their business operations \cite{Szczepanski2019EconomicAI}. Table III demonstrates the economic elements and data sources included in the development of the ranking.

\begin{table}[ht]
  \centering
  \caption{Educational elements for AI investment potential ranking}
    \begin{tabular}{p{5.39em}p{12.72em}p{9.835em}p{11.5em}}
    \toprule
    \multicolumn{4}{c}{EDUCATION/RESEARCH PILLAR} \\
    \midrule
    \multicolumn{1}{c}{Element} & \multicolumn{1}{c}{Description} & \multicolumn{1}{c}{Indicator} & \multicolumn{1}{c}{Source} \\
    \midrule
    \multicolumn{1}{c}{\multirow{3}[25]{*}{Education}} & \multicolumn{1}{c}{\multirow{3}[25]{12.72em}{Does the country have a sufficient population educated in order to implement and investigate AI solutions?}} & Population with tertiary education 25-34 years old (\%) & Population with tertiary education, OCDE \cite{OECDPopulationEducation} \\
\cmidrule{3-4}          &       & Population with tertiary education 55-64 years old (\%) & Population with tertiary education, OCDE \cite{OECDPopulationEducation} \\
\cmidrule{3-4}          &       & Number of Universities in Top 100 of Latin America & QS Latin America University Rankings 2023 \cite{QSQuacquarelliSymonds2023QS2023}\\
    \midrule
    \multicolumn{1}{c}{\multirow{3}[35]{*}{Research}} & \multicolumn{1}{c}{\multirow{3}[35]{12.72em}{What is the current status of research in the country to have the most up-to-date information regarding AI?}} & Articles Published in the AI Field & Emerging Technology Observatory, Country Activity Tracker (CAT): Artificial Intelligence \cite{CenterofSecurityandEmergingTechnology2023CountryIntelligence}\\
\cmidrule{3-4}          &       & Research and development expenditure (\% of GDP), 2019-2020 & World Bank \cite{WorldBankResearchCaribbean}\\
\cmidrule{3-4}          &       & Patents Granted in the AI Field & Emerging Technology Observatory, Country Activity Tracker (CAT): Artificial Intelligence \cite{CenterofSecurityandEmergingTechnology2023CountryIntelligence}\\
    \bottomrule
    \end{tabular}%
  \label{tab:education-elements}%
\end{table}%

\begin{longtable}[ht]{p{5.39em}p{12.72em}p{9.835em}p{11.5em}}
\caption{Economic Pillar Analysis} \label{tab:economic_pillar} \\

\toprule
\multicolumn{1}{c}{\textbf{Element}} & \multicolumn{1}{c}{\textbf{Description}} & \multicolumn{1}{c}{\textbf{Indicator}} & \multicolumn{1}{c}{\textbf{Source}} \\
\midrule
\endfirsthead

\multicolumn{4}{c}{{\tablename\ \thetable{} -- Continued from previous page}} \\
\toprule
\multicolumn{1}{c}{\textbf{Element}} & \multicolumn{1}{c}{\textbf{Description}} & \multicolumn{1}{c}{\textbf{Indicator}} & \multicolumn{1}{c}{\textbf{Source}} \\
\midrule
\endhead

\midrule
\multicolumn{4}{r}{{Continued on next page}} \\
\midrule
\endfoot

\bottomrule
\endlastfoot

\multirow{7}{*}{Startups} & \multirow{7}{12.72em}{What is the startup status in the country?} & Startups and other Privately-held AI companies headquartered in the selected country & Emerging Technology Observatory, Country Activity Tracker (CAT): Artificial Intelligence \cite{CenterofSecurityandEmergingTechnology2023CountryIntelligence}\\
\cmidrule{3-4} & & Number of unicorns in LATAM & CBINSIGHTS \cite{CBINSIGHTS2024TheCompanies}\\
\cmidrule{3-4} & & Startups by HQ (\%) & South Florida Journal of Development, Latin American Startups \cite{Campos2021LatinStartups}\\
\cmidrule{3-4} & & VC Investments in LATAM (\$ millions) 2024 & Dealroom.co \cite{Dealroom.co2024LatinAmerica}\\
\cmidrule{3-4} & & Startups By Sector (\%) & LAVCA, Latin American Startup Directory \cite{LAVCA2019LatinDIRECTORY}\\
\cmidrule{3-4} & & Startups with 1 or more women on the Executive Team (\%) & LAVCA, Gender Diversity in Latin American Tech/Startups \cite{LAVCA2019GenderTech/Startups}\\
\cmidrule{3-4} & & Startups with 1+ Female Investors in LATAM (\%) & LAVCA, Gender Diversity in Latin American Tech/Startups \cite{LAVCA2019GenderTech/Startups}\\
\midrule

\multirow{3}{*}{Salaries} & \multirow{3}{12.72em}{What is the cost of labor in the country?} & Median Salary for Computer Science roles & Glassdoor, Salaries \cite{Glassdoor2023Salary:2023}\\
\cmidrule{4-4} & & & Stack Overflow, 2022 Developer Survey \cite{StackOverflow20222022Survey}\\
\midrule

\multirow{9}{*}{Investments} & \multirow{9}{12.72em}{How much investment has the country received in the AI field?} & Estimated total value of incoming investments (millions USD) & Emerging Technology Observatory, Country Activity Tracker (CAT): Artificial Intelligence \cite{CenterofSecurityandEmergingTechnology2023CountryIntelligence}\\
\cmidrule{3-4} & & Incoming Investments & Emerging Technology Observatory, Country Activity Tracker (CAT): Artificial Intelligence \cite{CenterofSecurityandEmergingTechnology2023CountryIntelligence}\\
\cmidrule{3-4} & & Investment in Emerging Technologies & Network Readiness Index \cite{PortulansInstitute2023Network2023}\\
\cmidrule{3-4} & & Global Corporate Investment in AI by Investment Activity, 2013–22 (US Billion Dollars) & Stanford University, The AI Index Report \cite{StanfordUniversity2023Artificial2023}\\
\cmidrule{3-4} & & US\$4.6b LatAm Tech Invested across 440 transactions distribution 2019 & LAVCA's Annual Review of Tech Investment in Latin America \cite{LAVCA2020LAVCAsAmerica}\\
\cmidrule{3-4} & & US\$4.6b LatAm Tech Invested (DEALS) 2019 & LAVCA's Annual Review of Tech Investment in Latin America \cite{LAVCA2020LAVCAsAmerica}\\
\cmidrule{3-4} & & US\$4.09 LatAm Tech Invested across 488 transactions distribution 2020 & LAVCA's 2021 Review of Tech Investment in Latin America \cite{LAVCA2021LAVCAsAmerica}\\
\cmidrule{3-4} & & US\$4.09 LatAm Tech Invested (DEALS) 2020 & LAVCA's 2021 Review of Tech Investment in Latin America \cite{LAVCA2021LAVCAsAmerica}\\
\cmidrule{3-4} & & Total Capital Invested By Top Sectors 2020 & LAVCA's 2021 Review of Tech Investment in Latin America \cite{LAVCA2021LAVCAsAmerica}\\
\midrule

\multirow{3}{*}{Government} & \multirow{3}{12.72em}{How is the government involved with AI initiatives?} & Government promotion of investment in emerging technologies & University of Oxford \cite{PortulansInstitute2023Network2023}\\
\cmidrule{3-4} & & Number of AI-related bills passed into law by country, 2016-2023 & Stanford University \cite{StanfordUniversity2024Artificial2024}\\
\cmidrule{3-4} & & AI Government Strategy & Stanford University \cite{StanfordUniversity2024Artificial2024}\\
\midrule

\multirow{5}{*}{Economics} & \multirow{7}{12.72em}{What is the current and prospective economic status of the country?} & GDP 2023 & The World Bank, GDP (current US\$) - Latin America \& Caribbean \cite{WorldBank2023GDPCaribbean}\\
\cmidrule{3-4} & & GDP per capita, PPP, 2022 & The Global Economy, GDP per capita, PPP - Country rankings \cite{TheGlobalEconomyGDPRankings}\\
\cmidrule{3-4} & & Real GDP 2024 Projection (Annual Percent Change) & International Monetary Fund, World Economic Outlook \cite{InternationalMonetaryFund2023WorldRecovery}\\
\cmidrule{3-4} & & Real GDP 2028 Projection (Annual Percent Change) & International Monetary Fund, World Economic Outlook \cite{InternationalMonetaryFund2023WorldRecovery}\\
\cmidrule{3-4} & & Share of workers in "insulated GAI" classification by country & LinkedIn Economic Graph \cite{Baird2024GenerativeClassifications}\\
\midrule

\multirow{5}{*}{AI Adoption} & \multirow{5}{12.72em}{Are the Latin American countries implementing AI solutions into their processes?} & Organizations implementing AI & IDC, Analítica e Inteligencia Artificial para desarrollar una cultura basada en datos \cite{IDC2022AnaliticaDatos}\\
\cmidrule{3-4} & & AI Talent Concentration & Network Readiness Index \cite{PortulansInstitute2023Network2023}\\
\cmidrule{3-4} & & Adoption of Emerging Technologies & Network Readiness Index \cite{PortulansInstitute2023Network2023}\\
\cmidrule{3-4} & & Share of Respondents Who Say Their Organizations Have Adopted AI in at Least One Function, 2017–22 & McKinsey, The state of AI in 2022 \cite{Chui2022TheReviewb}\\
\cmidrule{3-4} & & AI Adoption by Organizations in LATAM between 2021 and 2022 & Stanford University, The AI Index Report \cite{Human-CenteredArtificialIntelligence2023Artificial2023}\\
\end{longtable}

\subsection{Elements and indicators}
Tables II, III, and IV summarize the elements and indicators used for the ranking.

\subsection{Missing values}
For countries with missing data, the ‘NA’ value was used to calculate the final scores (explained in the Calculating Scores section).

\subsection{Calculating scores}

\subsubsection{Normalization}
All scores were Z-score normalized using the following formula:

\begin{equation}
    z = (x -  \mu ) / ( \sigma )
\end{equation}

Where:

z is the normalized value

x is the original value

\( \mu \) is the media of the distribution

\( \sigma \) is the standard deviation of the distribution

All values used for normalization were original. Scores were calculated using the normalized formula based on the media of Latin American countries available data.

Z-score normalization is a method of standardization that transforms the values of a feature by subtracting the mean \( \mu \) and dividing by the standard deviation \( \sigma \). This results in a dataset with a mean of zero and a standard deviation of one.

This method was selected because it is less affected by outliers, which can skew the range and scaling of values in other methods. This method preserves the original distribution shape of values and allows for easy comparison regarding the number of standard deviations from the mean.

The value NA is commonly used to represent missing data. It acts as a placeholder for missing data and helps the formula recognize and handle the missing values correctly. NA stands for “not available” and can be used for both character and numeric data. Using NA allows the formula to ignore missing values during calculations or replace them with other methods.

\subsubsection{Limitations}
It is important to note that the data we included in our research is from 2019 to 2024, and the data from 2019 to 2020 may not accurately reflect the current situation in each country. Therefore, we urge caution when interpreting the results and when comparing them with more recent data from 2021 to 2024.

Additionally, we encountered challenges in obtaining data for some countries or regions, especially those with low internet penetration or political instability. As a result, our data may not be representative of the entire region, and we may have missed relevant aspects or trends related to our research question. To mitigate this issue, we used various data sources, listed in Table I, and cross-checked them for consistency and validity.

\subsubsection{Ranking}
The ranking was developed using indicators that allowed for the calculation of individual scores for each country—indicators that provided general information for all Latin American countries were not considered.

Implementing artificial intelligence correctly requires following a series of procedures while considering various factors such as infrastructure, education/research, and finances of the country or territory where it will be implemented. In Latin America, these factors can vary significantly in the region's countries, making Latin America attractive to investors by having many alternatives where AI can be implemented.

\section{RANKING LATAM COUNTRIES FOR INVESTMENT POTENTIAL BASED ON INFRASTRUCTURE, EDUCATION, AND ECONOMIC PILLARS}

Many factors make Latin America an attractive place for tech investments as mentioned before, these factors need to be carefully considered when valuing a startup, for this reason, we have developed a ranking of Latin American emerging powers for tech startup investments, with a focus on artificial intelligence (AI) as seen in Table IV. This ranking was developed considering three important pillars, infrastructure, education, and economics.

Infrastructure is a crucial factor when it comes to technology and AI deployment, it supports technological change and enables efficient strategy implementation \cite{Weiss1989TechnologicalStrategies}\cite{Amin2023ExploringImplementation}. To develop this ranking, many infrastructure factors were taken into account, including access to electricity and internet, internet penetration in rural and urban zones, internet speed for fixed broadband and mobile connections, availability of HPC systems in the region, availability of cloud services and number of providers, and notable ML models.

\begin{figure}[ht]
    \centering
    \includegraphics[width=.8\linewidth]{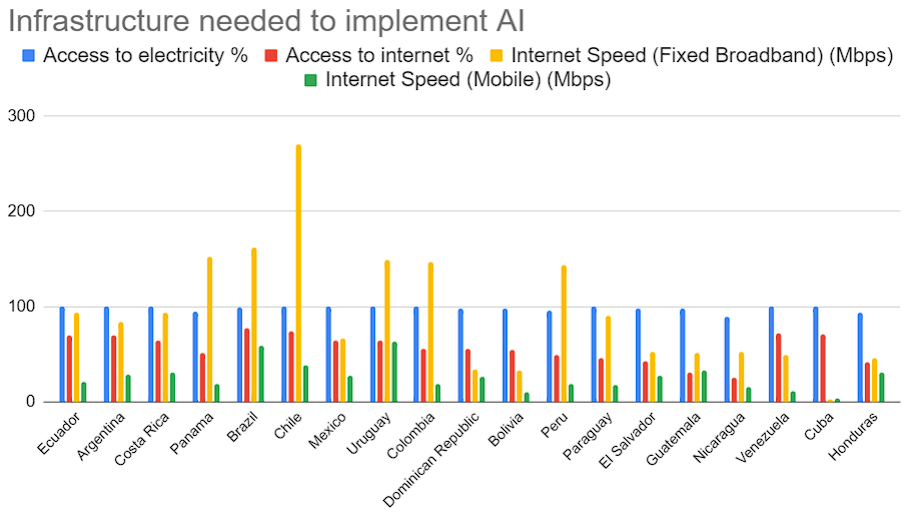}
    \caption{Comparing basic infrastructure elements needed to develop and implement AI in the Latin American region. Data for electricity access was sourced from the World Bank Group, access for internet access was sourced from the World Bank Group and the UNDP, and access for speed internet was sourced from the Speedtest Global Index.}
    \label{fig:Infra-chart}
\end{figure}

Access to electricity and internet services is a crucial requirement for any tech ecosystem. Reliable power ensures that businesses can operate continuously and efficiently. Fig \ref{fig:Infra-chart} shows that, for the Latin American region, most countries showed a great implementation of electricity systems, most of them reaching almost 100\%. Argentina and Ecuador, apart from the three current powers in LATAM (Brazil, Chile, and Mexico) reached 100\% access to electricity in their territories, in addition, Panama (95.3\%), Bolivia (97.6\%), and Peru (95.8\%) also show high levels of access, indicating a robust power infrastructure \cite{WorldBankOpenData2023AccessPopulation}.

\begin{table}[ht]
  \centering
  \caption{Ranking of Emerging Powers for Tech Investment Startups in Latin America}
    \begin{tabular}{p{5.39em}cccc}
    \toprule
    \textbf{Country} & \multicolumn{1}{c}{\textbf{Infrastructure Ranking}} & \multicolumn{1}{c}{\textbf{Education/Research Ranking}} & \multicolumn{1}{c}{\textbf{Finance Ranking
}} & \multicolumn{1}{c}{\textbf{AI Investment Ranking}} \\
    \midrule
    Brazil & 1     & 1     & 1     & 1 \\
    \midrule
    Mexico & 5     & 2     & 2     & 2 \\
    \midrule
    Chile & 2     & 3     & 3     & 3 \\
    \midrule
    Argentina & 3     & 4     & 6     & 4 \\
    \midrule
    Colombia & 6     & 5     & 4     & 5 \\
    \midrule
    Uruguay & 4     & 7     & 9     & 6 \\
    \midrule
    Costa Rica & 8     & 9     & 5     & 7 \\
    \midrule
    Panama & 11    & 15    & 7     & 8 \\
    \midrule
    Dominican Republic & \multirow{1}[4]{*}{15}    & \multirow{1}[4]{*}{12}    & \multirow{1}[4]{*}{8}     & \multirow{1}[4]{*}{9} \\
    \midrule
    Ecuador & 7     & 8     & 13    & 10 \\
    \midrule
    Peru  & 12    & 11    & 10    & 11 \\
    \midrule
    Venezuela & 9     & 10    & 15    & 12 \\
    \midrule
    Cuba  & 13    & 6     & 14    & 13 \\
    \midrule
    Honduras & 17    & 18    & 11    & 14 \\
    \midrule
    El Salvador & 10    & 13    & 18    & 15 \\
    \midrule
    Paraguay & 14    & 16    & 17    & 16 \\
    \midrule
    Guatemala & 18    & 19    & 12    & 17 \\
    \midrule
    Bolivia & 16    & 17    & 19    & 18 \\
    \midrule
    Nicaragua & 19    & 14    & 16    & 19 \\
    \bottomrule
    \end{tabular}%
  \label{tab:investment-ranking}%
\end{table}%

Internet access is essential for connecting people and businesses to digital services, cloud platforms, online markets, and others. It is also important to measure internet penetration in urban and rural areas to have a complete understanding of this aspect in Latin American countries. According to the United Nations Development Program, Brazil and Venezuela were the top performers in this factor, with 77\% and 72\% of internet access respectively. Other emerging powers such as Argentina (70\%) and Costa Rica (65\%) show potential for internet access growth. On the other hand, the emerging powers Ecuador (60\% rural penetration), Panamá (Urban 61\%, Rural 61\%), and Argentina (Urban 78\%, Rural 38\%) demonstrate a moderate urban and rural reach \cite{UnitedNationsDevelopmentProgrammeUNDPInternetCaribbean}.

\begin{figure}[ht]
    \centering
    \includegraphics[width=.8\linewidth]{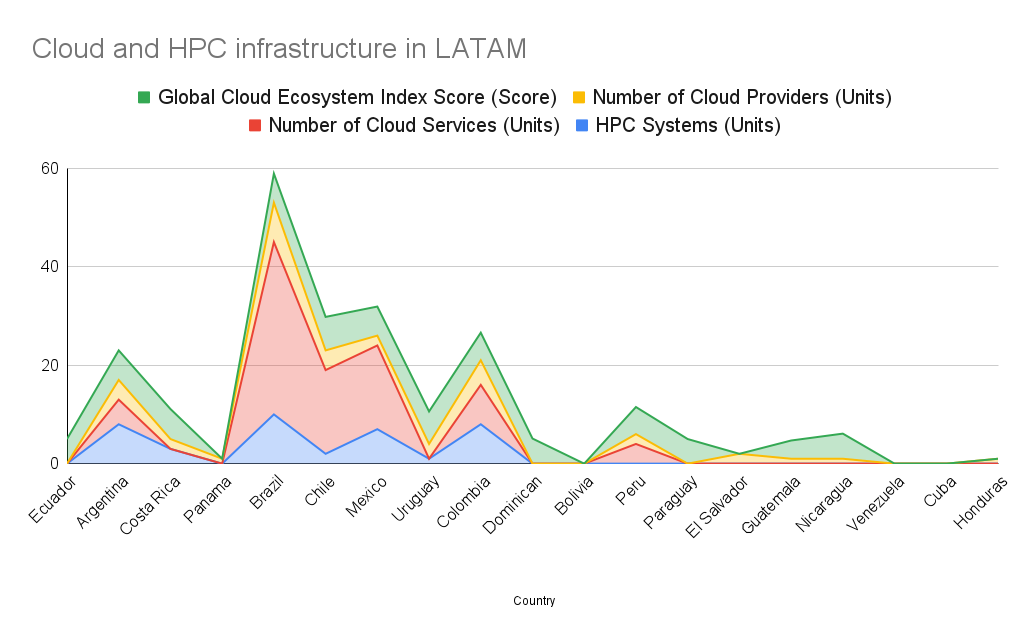}
    \caption{Cloud ecosystem elements needed for AI development and implementation. The Global Cloud Ecosystem Index was sourced from MIT, the number of cloud ecosystems was sourced from Tele Geography, the number of cloud services was sourced from Data Center Map, and HPC systems distribution in LATAM was sourced from RISC2.}
    \label{fig:cloud-chart}
\end{figure}

Latin America must have High-Performance Computing Systems (HPC Systems) in place to facilitate the development and implementation of AI. These systems are essential for advanced AI research, large-scale data processing, and running complex AI algorithms. As shown in Fig \ref{fig:cloud-chart}, Brazil currently leads with 10 HPC systems, while emerging powers like Argentina, Colombia, and Costa Rica have a notable HPC infrastructure with 8, 8, and 3 HPC systems respectively. Uruguay and Chile also have 1 and 2 HPC systems, indicating a growing capacity for advanced computational tasks \cite{Hafner2021RISC2LATAM}.

Fig \ref{fig:cloud-chart} also shows the number of cloud services and providers, which is a critical factor in determining the robustness of a tech ecosystem. Brazil leads with 35 cloud services and 8 providers, indicating a well-established infrastructure. Argentina also shows competitive strength with 5 services and 4 providers, as well as Colombia with 8 services and 5 providers, highlighting their growing cloud ecosystem \cite{TeleGeography2024CloudMap}\cite{DataCenterMapCloudInfrastructure}.

In terms of AI research, Argentina stands out with 80.5 notable machine learning models developed between 2003 and 2023, showcasing its significant R\&D capabilities \cite{StanfordUniversity2024Artificial2024}.

The Global Cloud Ecosystem Index Score, shown in Fig \ref{fig:cloud-chart}, indicates the preparedness of countries to support AI infrastructure. Chile has the highest score at 6.80, followed by Uruguay at 6.60 and Argentina at 6.00 \cite{MITTechnologyReview2022Global2022}. These scores reflect the strong potential for emerging powers in AI investment. Overall, these metrics highlight the promising landscape for AI investments in Latin America.

The education pillar is crucial as it indicates the foundational aspects of education that can lead to the development of a highly skilled workforce necessary for the advancement and growth of the AI and tech industries. By focusing on developing a workforce with strong cognitive and technical skills, the education pillar plays a vital role in shaping the future of AI and technology, enabling innovation, and driving economic progress.

Tertiary education is one of the important aspects taken into consideration in this pillar, divided into two categories, young adults and old adults. Chile stands out with the highest percentage of young adults with tertiary education (40.5\%), which reflects the strong commitment to higher education among its youth. The emerging powers, Colombia (27.1\%) and Argentina (19\%), also demonstrated progress in this aspect regarding young people. For older adults, the emerging power Argentina leads with 24.9\%, suggesting a long-standing tradition of higher education and the possibility of the older generation being involved actively and contributing to the economy of the country. Argentina is followed by two of the three current powers in the AI sector, Brazil (17.1\%) and Chile (18.9\%) \cite{OECDPopulationEducation}.

\begin{figure}[ht]
    \centering
    \includegraphics[width=.65\linewidth]{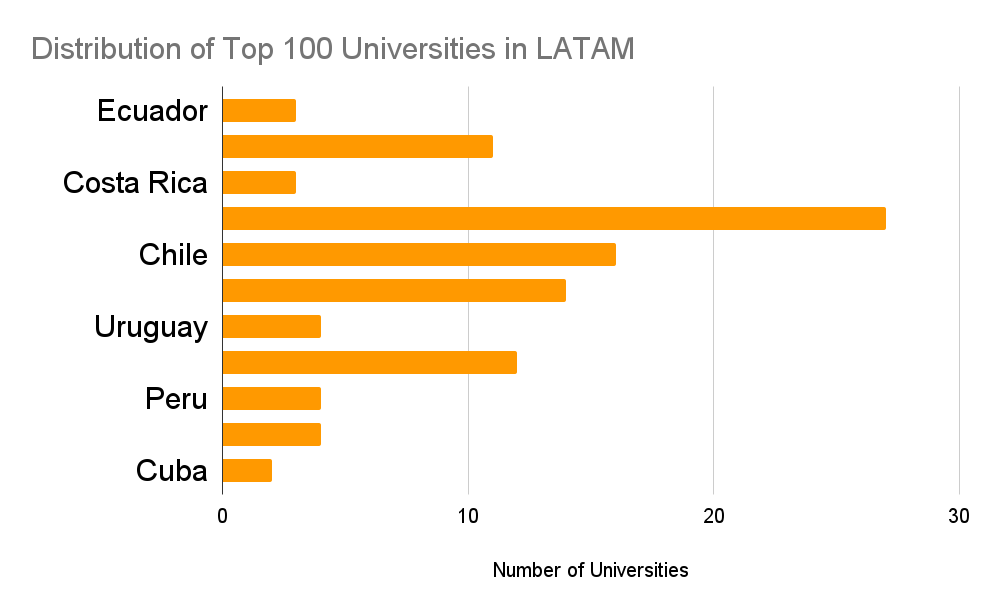}
    \caption{Distribution of the TOP 100 universities in Latin America. Data was sourced from the QS World University Ranking.}
    \label{fig:Universities-chart}
\end{figure}

Additionally, Fig \ref{fig:Universities-chart} shows how Brazil dominates the academic institutions' quality and research capabilities with 27 universities in the top 100 of Latin America, the emerging powers Colombia (12) and Argentina (11) also performed well, indicating great education systems established in Latin America \cite{QSQuacquarelliSymonds2023QS2023}.

\begin{figure}[ht]
    \centering
    \includegraphics[width=.8\linewidth]{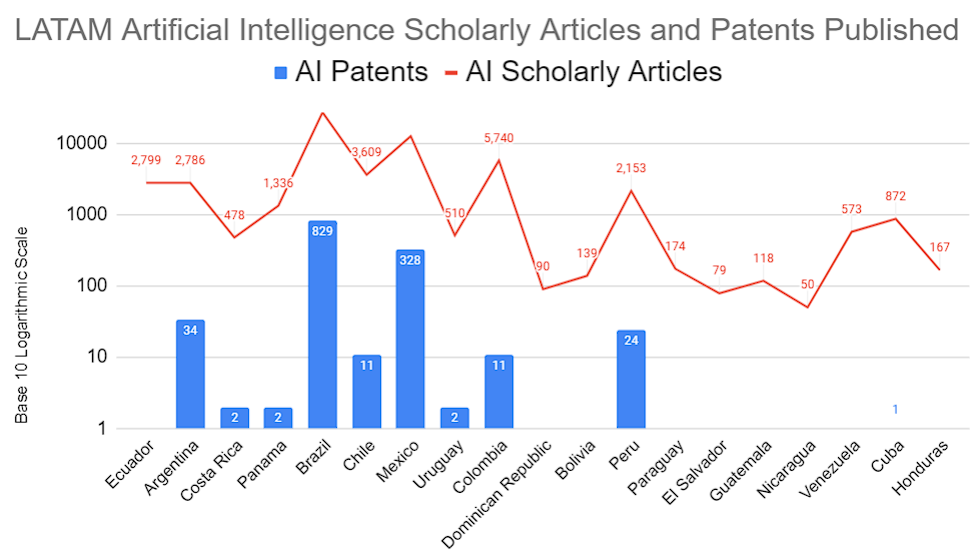}
    \caption{Comparing academic levels in LATAM through the research and patents status in the region. Data was sourced from the Emerging Technology Observatory.}
    \label{fig:RandP-chart}
\end{figure}

The academic level in Latin America can be observed through the number of articles and patents published in the field of AI. Fig \ref{fig:RandP-chart} Brazil leads by a significant margin, with 27,140 publications and 829 patents, reflecting its active research and innovative contributions to the field. Other emerging powers have also increased their levels of research and applications in the AI sector, such as Ecuador with 2,799 publications, Argentina with 2,786 publications and 34 grants, Colombia with 5,740 publications and 11 grants, and Peru with 2,153 publications and 24 grants \cite{CenterofSecurityandEmergingTechnology2023CountryIntelligence}.

\begin{figure}[ht]
    \centering
    \includegraphics[width=.65\linewidth]{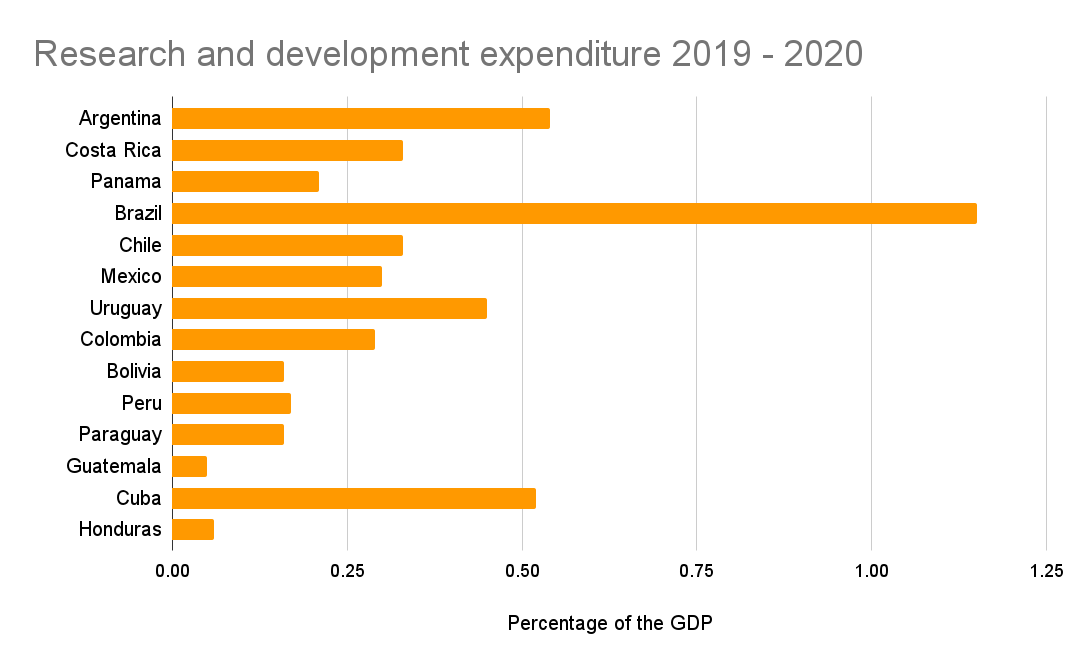}
    \caption{Research and development expenditure made in Latin America for the years 2019-2020. Data was sourced from the World Bank Group.}
    \label{fig:RandD-chart}
\end{figure}

The research and development expenditure also plays an important role in the educational pillar. In this case, Fig \ref{fig:RandD-chart} shows that Brazil allocates 1.15\% of its GDP to R\&D, being the highest among the Latin American countries, which can be noted in their articles, patents, and educational institutions in its territory. Argentina (0.54\%) and Chile (0.33\%) also invest in R\&D, reflecting their commitment to supporting scientific research and technological innovation \cite{WorldBankResearchCaribbean}.

Looking at the third pillar, economic status is crucial for measuring not only the potential of each specific country to implement AI but also for analyzing the macroeconomic factors that can directly or indirectly impact startups in a specific country.

It is important to highlight the entrepreneurial activity and the presence of AI-driven businesses as critical indicators of a country's potential in the AI sector. Currently, Brazil leads this activity with 348 startups, followed by Mexico with 83 startups, and Chile with 79 startups. However, emerging powers are also growing their entrepreneurial ecosystem, showing strong potential for economic growth. Colombia has a notable number of startups with 37, while Argentina reflects a vibrant entrepreneurial ecosystem with 34 startups. Peru and Uruguay show promise with a moderate number of startups, 14 and 7 respectively. Additionally, El Salvador is in the early stages of developing its startup ecosystem with 3 startups, and despite economic challenges, other emerging powers like Venezuela also have startup activity, with 2 startups \cite{CenterofSecurityandEmergingTechnology2023CountryIntelligence}.

\begin{figure}[ht]
    \centering
    \includegraphics[width=.7\linewidth]{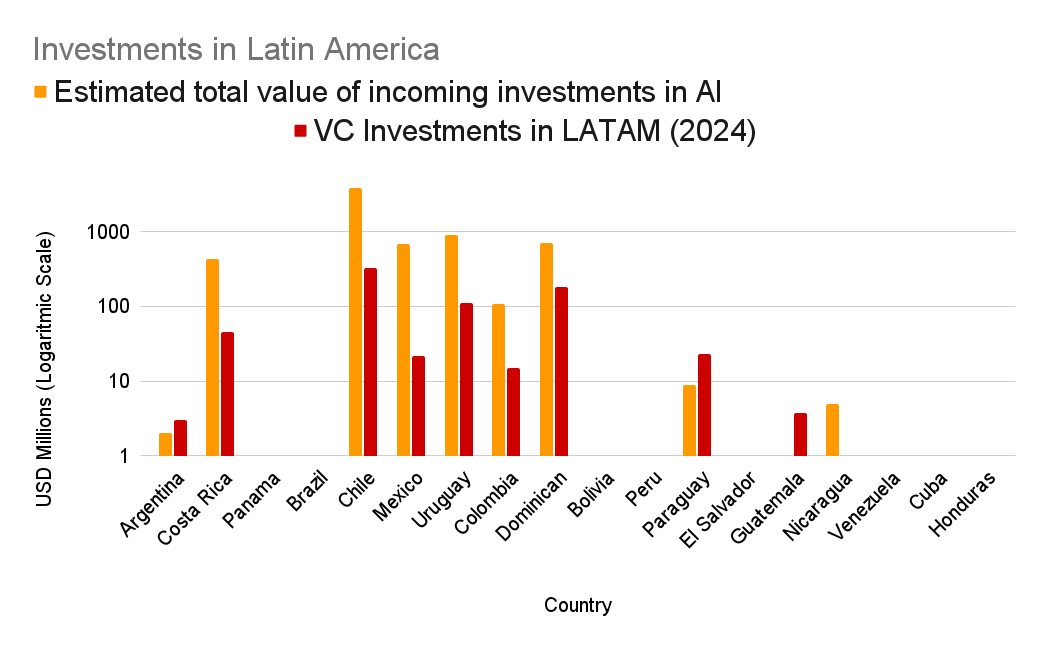}
    \caption{Investment status in the technology and AI sector in Latin America. Data was sourced from Dealroom.co and The Emerging Technology Observatory.}
    \label{fig:Ai-investments-chart}
\end{figure}

The investment factor also provides an understanding of financial support and investor confidence in the AI sector of each country. When it comes to the estimated value of incoming VC investments apart from the three powerhouse countries in Latin America, Fig \ref{fig:Ai-investments-chart} shows the emerging powers such as Colombia (\$697 million), Argentina (\$431 million), and Uruguay (\$109 million), which demonstrated a significant investment in the AI sector. These high levels of investment reflect strong confidence in the countries' tech potential and growth prospects. Other emerging powers are also in the early stages of development due to economic instability and other factors but they still present an opportunity. For example, Peru (\$9 million), Guatemala (\$5 million), Ecuador (\$2 million), El Salvador (\$1 million), Panama (\$1 million), and Venezuela (\$1 million) \cite{CenterofSecurityandEmergingTechnology2023CountryIntelligence}.

In addition, venture capital (VC) investments in general, not only in the AI sector, indicate the growth potential of the ecosystem and the interest and support from the investment community. Significant VC investments indicate strong growth potential. For example, in 2024, Fig \ref{fig:Ai-investments-chart} shows that there have been substantial investments in Latin America, as seen when analyzing countries like Colombia (\$183.9 million), Argentina (\$45.3 million), Peru (\$22.8 million), Uruguay (\$14.9 million), El Salvador (\$3.8 million), and Ecuador (\$3 million) \cite{Dealroom.co2024LatinAmerica}.

\begin{figure}[ht]
    \centering
    \includegraphics[width=.7\linewidth]{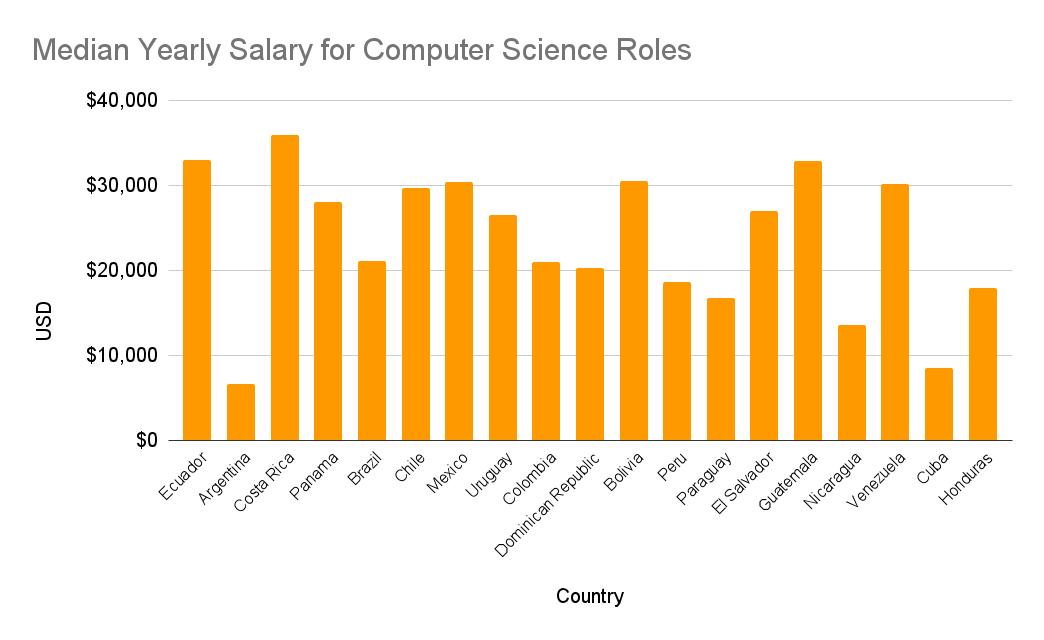}
    \caption{Comparing the salary in Latin America for roles in the Computer Science field. Data was sourced from Glassdoor and Stack Overflow.}
    \label{fig:Salaries-chart}
\end{figure}

Before making an investment decision, it is important to consider factors such as the cost of the investment. In the field of artificial intelligence (AI), understanding the cost of developing and implementing AI is crucial. One significant aspect to consider is the yearly salary for computer scientists, which is shown in Fig \ref{fig:Salaries-chart}, who will be responsible for developing and implementing artificial intelligence in operational processes. These salaries vary significantly across countries. In Latin America, the countries with the lowest costs to develop AI are Argentina (\$6,500-23,600), Venezuela (\$17,000-30,000), Colombia (\$20,900-23,200), Paraguay (\$11,200-16,700), Peru (\$18,500-36,700), and Cuba (\$4,200-8,500) as compared to the US, where the median salary is \$104,420 annually \cite{Glassdoor2023Salary:2023}\cite{StackOverflow20222022Survey}\cite{USBureauofLaborStatistics2023SoftwareHandbookb}.

The government strategies focused on AI are important for measuring national priorities, and policies, and creating a supportive environment for AI development. It also involves providing a regulatory framework to ensure the ethical use of AI. In this sense, the governments of emerging powers such as Argentina, Uruguay, Colombia, Peru, and Cuba have already implemented AI strategies, indicating strong support for AI development. Other emerging powers, such as Costa Rica and Venezuela, have also initiated the process of developing their strategies \cite{StanfordUniversity2024Artificial2024}.

The promotion of government investment in emerging technologies signals to investors that a country is supportive of innovation and open to foreign investment. According to the Network Readiness Index 2023, Uruguay (43.54), Chile (38.06), Mexico (36.8), and Colombia (37.92) are leading the region in this aspect, showing strong performance. However, there is still room for improvement \cite{PortulansInstitute2023Network2023}.

When it comes to tech investments, high levels of this type of investment indicate the availability of startups and tech companies, which is crucial for scaling and growth. It also reflects market confidence in a specific country's tech sector. In 2019 and 2020, the three major AI powers in LATAM received the highest levels of tech investments. Brazil had the highest levels of tech investments, receiving \$2,490 million in 2019 and \$2,380 million in 2020. Argentina and Colombia also saw significant investments during these years. Argentina received \$290 million in 2019 and \$222 million in 2020, while Colombia received \$1,090 million in 2019 and \$469 million in 2020 \cite{LAVCA2020LAVCAsAmerica}\cite{LAVCA2021LAVCAsAmerica}.

These metrics, together with the expected growth of the Latin American countries provide a better understanding of the economic status of the region, making an integral view with the analysis of the three pillars in conjunction. Brazil, Mexico, and Argentina had the largest economies showing the highest GDP in 2023, with Brazil at \$2.2 trillion, Mexico at \$1.8 trillion, and Argentina at \$640 billion \cite{WorldBank2023GDPCaribbean}. However, other emerging powers in LATAM showed great predictions in the percentual change of the Real GDP for 2024 and 2028, being Costa Rica (+3.2\% for 2024 and 2028), Venezuela (+4.5\% for 2024), Guatemala (+3.5\% for 2024 and +3.9\% for 2028), Honduras (+3.5\% for 2024 and +3.9\% for 2028), and Paraguay (+3.5\% for 2024 and 2028) the countries with the best-projected growth \cite{InternationalMonetaryFund2023WorldRecovery}.

\section{RISK MANAGEMENT IN AI INVESTMENTS}

The growth of the tech startup ecosystem in Latin America (LATAM) is propelled by innovative entrepreneurs who are addressing market needs across various sectors. However, startups encounter unique risks when introducing new technologies, requiring specific risk management approaches \cite{Teberga2018IdentificationStart-ups}.

Political risk in Latin America presents significant challenges for startups and investors. The region is known for recurring political risks stemming from entrenched power groups \cite{Salamanca2016PoliticalAmerica}. This often leads to macroeconomic turbulence, excessive bureaucracy, political instability, corruption, and violence \cite{Jones2018Business1970}. These factors contribute to higher sovereign spreads, especially when coupled with a weak rule of law or low-quality regulation \cite{Hansen2016PoliticalAmerica}. For venture capital investments, political hazards can reduce investment size, while low-quality legal systems are related to larger investments but disrupt traditional VC-staging strategies \cite{Khoury2015NavigatingQuality}. The political landscape in Latin America requires investors to carefully assess risks, as current forecasting methods may not be sufficient to comprehend the full range of potential issues \cite{Salamanca2016PoliticalAmerica}. Understanding these political risk factors is crucial for startups and investors looking to navigate the complex business environment in Latin American countries.

The tech startup ecosystem is growing rapidly due to the potential of the Latin American region, where its ecosystem value has increased by millions of dollars \cite{Pena2021TecnolatinasAge}. Fintech startups are revolutionizing the financial industry, with companies like Nubank (Brazil) and Rappi (Colombia) emerging as unicorns and receiving substantial international funding \cite{Campos2021LatinStartups}. However, this can augment the competitive landscape, which can present several risks. Market saturation and increased competition in emerging markets can lead to challenges for businesses. Copycat competitors from emerging economies often develop less expensive products, potentially diluting the market presence of established firms \cite{YadongLuo2011EmergingStrategy.}. High-growth markets also pose risks, including price pressures and the need for substantial investments \cite{Aaker1986TheMarkets}. Companies must carefully weigh the benefits against the costs and consider the unique challenges of each market before entry \cite{Khanna2005StrategiesMarkets.}\cite{Aaker1986TheMarkets}.

The issue of tax incentives for promoting innovation and investment in Latin America has been extensively studied and debated. These incentives aim to stimulate economic growth; however, their effectiveness and efficiency are often questioned \cite{Roca2010EvaluationBenefits}. In comparison to nations with similar GDP per capita, Latin American countries generally exhibit lower performance in innovation indicators, particularly in business R\&D and university contributions to innovation \cite{Hall2005HallAmerica}. The implementation of tax incentives in the region is challenging due to bureaucratic hurdles and administrative costs \cite{Roca2010EvaluationBenefits}. An analysis of historical approaches to tax incentive legislation in seven Latin American countries reveals that policy decisions are influenced by public perception and cost-benefit evaluations \cite{Byrne2002TaxCountries}. Despite these efforts, Latin American countries may struggle to compete globally due to TRIPS agreement requirements and patent policies \cite{Hall2005HallAmerica}. Additionally, the effectiveness of tax incentives in Latin America is complicated by the need to adapt policies to local economic conditions rather than simply adopting U.S. models \cite{FROOMKIN1957SOMEAMERICA}. Compared to other major countries such as China and the US, where there are different tax incentives for startups and R\&D investments, Latin America still needs to implement these types of incentives to generate more international interest from investors.

Economic instability in Latin America presents significant risks for startups and businesses. The higher country risk, covering economic, financial, and political factors, leads to increased stock market volatility in the region \cite{Cermeno2024CountryAmerica}. Startups introducing new technologies need to adopt a specific approach to risk management, taking into account factors such as innovation, digital solutions, and scalability \cite{Teberga2018IdentificationStart-ups}. Entrepreneurial risk attitudes in Latin America are influenced by gender, education, skills, age, and prior business experience \cite{EspinosaSepulveda2014LosVigentes}. While Latin American economies have made progress in addressing traditional macroeconomic issues, structural sources of volatility persist, and new challenges have emerged in the improved economic environment \cite{Caballero2000MacroeconomicStudies}. These factors collectively contribute to a complex risk landscape for startups in the region. Therefore, careful consideration of economic volatility and its potential impacts on business development and success is necessary.

To mitigate these risks, it is extremely important to take a comprehensive approach to the LATAM market. Addressing and evaluating each of the startups' inherent risks will provide a better understanding of the region, which will make the investor adopt specific risk management strategies.

\section{FRAMEWORK FOR VALUING STARTUPS IN LATAM AND FOOD DELIVERY STARTUP CASE STUDY
}

In this section, we will explore the TAM, SAM, and SOM metrics for the online food industry. These metrics can help us estimate the industry's growth potential and market profitability and serve as an example for investing in artificial intelligence in Latin America using the DCF method to value a startup with similar estimated cash flows.

The TAM, SAM, and SOM are different metrics that are used together to understand how far an organization can reach within a specific market. The TAM (Total Addressable Market) measures the total market demand for a specific product, which is the maximum amount of demand the product can achieve. The SAM (Serviceable Available Market) represents the portion of the TAM that the product can realistically be sold to. Finally, the SOM (Serviceable Obtainable Market) indicates the actual demand that the product can obtain, considering competition, strategy, and market share. The DCF method is a comprehensive valuation approach that calculates the present value of expected future cash flows.

To calculate the Total Addressable Market (TAM) of the studied countries, we considered the monthly price of the basic food basket in both urban and rural areas. After excluding the population under 14 years old, we factored in the total population and the percentage of people living in urban and rural areas. The total TAM for the food delivery industry was then calculated by multiplying the percentage of the different zones with the cost of the basic food basket and then multiplying it by 12 to calculate the annual TAM.

To calculate the Serviceable Available Market (SAM), we excluded certain population groups that may have difficulty using a food delivery service because of limited access to technology. Specifically, we excluded the visually impaired population, the extremely poor, individuals over 64 years old, and those without internet access from the study population (some exclusions were not included due to unavailability or outdated data.). We then multiplied the remaining population by the total cost of the basic food basket.

In calculating the Serviceable Obtainable Market (SOM), we took into account various market shares based on market challenges, competency, economic factors, and other relevant considerations. The market shares of the countries under study ranged from 5\% to 20\%. The SOM was determined by multiplying the Serviceable Addressable Market (SAM) by the total market share and then projecting the actual SOM over the next 5 years using a growth rate for the SAM ranging from 0.5\% to 1.4\% based on the potential increase of the countries being studied.

By analyzing all Latin American countries and adapting the TAM, SAM, and SOM to the use of artificial intelligence in the online food service industry, we can highlight that Mexico, Colombia, and Brazil are the countries that have the highest SOM, with \$19 mil, \$16 mil, and \$16 mil respectively. Tables V, VI, and VII demonstrate the calculations for Mexico as a reference. This means that for an online food sales startup that applies artificial intelligence, considering the competition and the strategies used, these quantities demonstrate the magnitude of maximum demand that they can obtain prospectively by the year 2025, with an average increase with respect to the following years of between 0.5\% to 1.4\% depending on the country.

Likewise, it is also highlighted that for almost all the countries for which the TAM, SAM, and SOM were calculated taking into account prices of the basic shopping basket, number population, percentages of poverty and population without access to the internet, and the age distribution in the population, the majority had positive numbers, thus estimating profitability for startups created in Latin America. However, countries such as Guatemala, Nicaragua, Venezuela, Cuba, and Honduras obtained negative results, implying a lack of profitability for startups opened in these countries that are within the food industry and applying artificial intelligence. This is due to various factors, since in these countries the levels of poverty and population without internet access are very high, causing them not to be considered as potential applicants for this type of services since they would be more likely to resort to more traditional services, such as going directly to a supermarket. It should be noted that this does not demonstrate an absolute truth about the negative profitability that a startup may have within these countries, since these analyses are prospective and based on the assumption of facts, so it cannot be confirmed.

\begin{table}[ht]
  \centering
  \caption{TAM}
    \begin{tabular}{lrrr}
    \toprule
    \multicolumn{4}{c}{Mexico} \\
    \midrule
    \multicolumn{4}{c}{Food Industry TAM} \\
    \midrule
    “Basic Food Basket” monthly per person: &       &       &  \\
    Urban Zones (MXN) &       &       & 1,644.00 \\
    Rural Zones (MXN) &       &       & 2,144.00 \\
    Pupulation in México 2021 &       &       & 126,705,138.00 \\
    Assuming the basic diet prices start with people more than 15 years old: &       &       &  \\
    Population < 14 yo &       &       & 24.95\% \\
    Population > 14 yo &       &       & 75.05\% \\
    Population number > 14 yo &       &       & 95,092,206.07 \\
    Population Urban \% &       &       & 79\% \\
    Population Number for Urban Zones &       &       & 75,122,842.79 \\
    Population Rural \% &       &       & 21\% \\
    Population Number for Rural Zones &       &       & 19,969,363.27 \\
    \midrule
    FOOD INDUSTRY TAM TOTAL IN MXN &       &       & 1,995,795,220,976.17 \\
    \midrule
    FOOD INDUSTRY TAM TOTAL IN USD &       &       & 106,979,373,836.95 \\
    \bottomrule
    \end{tabular}%
  \label{tab:tam-table}%
\end{table}%

\begin{table}[ht]
  \centering
  \caption{SAM}
    \begin{tabular}{lrrr}
    \toprule
    \multicolumn{4}{c}{Mexico} \\
    \midrule
    \multicolumn{4}{c}{Food Industry SAM Adapted to AI} \\
    \midrule
    FOOD INDUSTRY TAM TOTAL IN USD &       &       & 106,989,592,308.48 \\
    Blind Population &       &       & 415,800.00 \\
    Extreme poverty &       &       & 10,900,000.00 \\
    Population > 64 yo &       & 8.13\% & 10,301,127.72 \\
    Pop withouth internet connection &       & 31.50\% & 39,912,118.47 \\
    Population < 14 yo &       &       & 31,612,931.93 \\
    Prorate "Basic Food Basket" monthly in MXN &       &       & 1,749.00 \\
    \midrule
    FOOD INSUTRY SAM IN MXN &       &       & 704,423,599,553.05 \\
    \midrule
    FOOD INDUSTRY SAM IN USD &       &       & 37,762,388,113.04 \\
    \bottomrule
    \end{tabular}%
  \label{tab:sam-table}%
\end{table}%

In our analysis of estimated values in the food industry across different countries using artificial intelligence and based on different estimated cash flows using TAM, SAM, and SOM, we used Brazil as the reference country, which ranked 1st in the ranking. Brazil is the leading country in Latin America with the highest levels of education, infrastructure, and economics. Our analysis suggests that a startup in Brazil could potentially generate an average of \$890,000 annually from 2025 to 2029. Considering the inherent risks, the present value of that startup is estimated at \$2.5 million, as shown in Table VIII for reference.

\begin{table}[ht]
  \centering
  \caption{SOM}
    \begin{tabular}{rrrr}
    \toprule
    \multicolumn{4}{c}{Mexico} \\
    \midrule
    \multicolumn{4}{c}{Food Industry SOM Adapted to AI} \\
    \midrule
    \multicolumn{2}{p{10.78em}}{Market Share} & 10\%  &  \\
    \multicolumn{2}{p{10.78em}}{SOM} & 3,777,577,216.14 &  \\
          &       &       &  \\
    2025  &       & 18,887,886.08 & 0.50\% \\
    2026  &       & 20,776,674.69 & 0.55\% \\
    2027  &       & 22,854,342.16 & 0.61\% \\
    2028  &       & 25,139,776.37 & 0.67\% \\
    2029  &       & 27,653,754.01 & 0.73\% \\
    \bottomrule
    \end{tabular}%
  \label{tab:som-table}%
\end{table}%

Interestingly, our analysis shows that Colombia, as an emerging power, has surpassed these estimations. Based on the same calculations adapted to the studded country conditions, the estimated average cash flows for a startup in Colombia were \$1 million, with the present value of that startup estimated at \$2.65 million.

Ecuador is the second emerging market with the best valuation of a food service startup using AI, with a present value of \$368k, followed by Argentina with \$131k, Dominican Republic with \$113k, Peru with \$113k, and Costa Rica with \$107k.

\begin{table}[ht]
  \centering
  \caption{Startup Valuation}
    \begin{tabular}{c|c|c|c|c|c}
    \toprule
    \multicolumn{6}{c}{Brazil} \\
    \midrule
    \multicolumn{6}{c}{Estimated Value of a Startup} \\
    \midrule
    Year & Revenue & COGS & Gross Profit & Operating Expenses & Operating Income \\
    \midrule
    2025  & 1,666,519.87 & 416,629.97 & 1,249,889.90 & 374,966.97 & 874,922.93 \\
    \midrule
    2026  & 1,979,825.60 & 494,956.40 & 1,484,869.20 & 445,460.76 & 1,039,408.44 \\
    \midrule
    2027  & 2,332,594.53 & 583,148.63 & 1,749,445.90 & 524,833.77 & 1,224,612.13 \\
    \midrule
    2028  & 2,729,135.60 & 682,283.90 & 2,046,851.70 & 614,055.51 & 1,432,796.19 \\
    \midrule
    2029  & 3,400,922.82 & 850,230.71 & 2,550,692.12 & 765,207.64 & 1,785,484.48 \\
    \midrule
    Taxes & Cash Flows & Present Value Of Cash Flows & Terminal Value & \multicolumn{1}{p{5.39em}|}{Present Value\newline{}Of Terminal Value} & Total Present Value \\
    \midrule
    262,476.88 & 612,446.05 & 453,663.74 & \multirow{5}[10]{*}{3,387,721.87} & \multirow{5}[10]{*}{755,507.72} & \multirow{5}[10]{*}{2,537,498.90} \\
\cmidrule{1-3}    311,822.53 & 727,585.91 & 399,224.09 &       &       &  \\
\cmidrule{1-3}    367,383.64 & 857,228.49 & 348,413.75 &       &       &  \\
\cmidrule{1-3}    429,838.86 & 1,002,957.33 & 301,958.59 &       &       &  \\
\cmidrule{1-3}    535,645.34 & 1,249,839.14 & 278,731.00 &       &       &  \\
    \bottomrule
    \end{tabular}%
  \label{tab:valuation-table}%
\end{table}%

\subsection{Startup estimated evaluation}
In this section, we will outline the methodology used to calculate the valuation of startups in the online food industry in Latin America, focusing on key financial metrics and applying the Discounted Cash Flow (DCF) method to determine the present value of these startups. 

\subsection{Revenue calculation}
The revenue for each year from 2025 to 2029 was calculated based on 10\% of the Serviceable Obtainable Market (SOM) for their respective years. For subsequent years, an additional 1\% was added to the revenue. This incremental approach allows us to estimate the potential revenue growth over the specified period.

\subsection{Cost of goods sold (COGS)}
The COGS was assumed to be 25\% of the total revenue for each year. This percentage was consistently applied across all years to maintain a standard margin for direct costs associated with the products or services sold taking into account that the SaaS companies see an average COGS in the range of 5\% to 40\% \cite{ChurnzeroCostCOGS}.

\subsection{Operating expenses}
Operating expenses were estimated at 30\% of the total Gross Profit for each year. This assumption accounts for the expenses required to run the business operations, excluding the costs directly tied to the production of the software.

\subsection{Taxes}
Taxes were calculated as 30\% of the total Operating Income for each year, based on the media of the income tax in Latin America \cite{OCDE2024Estadisticas2024}. This represents the income tax rate applied to the Operating Income.

\subsection{Cash flows}
Cash Flows were calculated by subtracting the taxes from the Operating Income. This represents the net cash generated from operating activities after accounting for taxes.

\subsection{Present value of cash flows}
The Present Value of Cash Flows for the years 2025 to 2029 was calculated using the DCF formula, applying different discount rates based on the ranking developed. The discount rates were distributed as follows:

Countries Ranked 1-3: 35\%
Countries Ranked 4-6: 38\%
Countries Ranked 7-10: 40\%
Countries Ranked 11-13: 42\%
Countries Ranked 14-16: 45\%
Countries Ranked 17-19: 50\%

These rates reflect the varying levels of risk associated with investing in different Latin American countries, where the highest-ranked countries represent less risk, and the lowest-ranked countries represent more risk.

\subsection{Terminal value and present value of terminal value}
The Terminal Value was calculated to estimate the value of the startup beyond the forecast period (2025-2029). The Present Value of Terminal Value was then determined using the respective discount rates.

\subsection{Total present value}
Finally, the Total Present Value was obtained by summing the Present Value of Cash Flows from 2025 to 2029 and the Present Value of Terminal Value. This total provides a comprehensive valuation of the startup, accounting for both the projected cash flows and the long-term value beyond the forecast period.

This methodology was applied uniformly across the analysis, allowing for a consistent comparison of startup valuations across different Latin American countries. The results highlight the potential profitability and investment attractiveness of startups in the online food industry using artificial intelligence, with specific insights into the leading and emerging markets within the region.

\section{CONCLUSION}
Latin America has great potential for development, which can be enhanced by foreign investment and the strategic implementation of artificial intelligence. Investing in AI can greatly benefit various sectors of the region's economy by reducing costs, increasing productivity, and promoting technological advancements.

The imbalance of indicators we have investigated can increase investment risk and make the implementation of artificial intelligence more complex. Therefore, countries need to focus on improving their infrastructure, academic status, and economic conditions to become potential destinations for investment and the implementation of artificial intelligence.

The valuation of technology startups in Latin America presents significant opportunities and challenges. Traditional valuation methods alone are insufficient due to the unique characteristics and dynamic nature of tech startups, especially those leveraging artificial intelligence. By adopting a comprehensive approach that incorporates metrics like TAM, SAM, and SOM, and utilizing the Discounted Cash Flow (DCF) method, we can better understand and estimate the potential profitability and market viability of these startups.

In summary, the emergence of several Latin American countries as potential powers for implementing artificial intelligence and receiving investment is a positive development.

The Latin American region has already been receiving foreign investment and implementing strategies for the proper use of artificial intelligence technology. However, compared to other regions of the world, the adoption of these technologies is still low. Therefore, it is crucial to continue promoting investment and adoption of artificial intelligence to strengthen initiatives and collaborations in this field, contributing to the growth of the countries' economies.

Latin America has the potential to advance its development through the adoption of artificial intelligence. Therefore, governments, companies, and universities should collaborate to create an ecosystem conducive to technological innovation, adoption, and investment.

Investing in startups within these emerging markets offers a strategic advantage in portfolio diversification. The high growth potential and evolving market landscapes present unique opportunities for significant returns. By spreading investments across different countries and sectors, investors can mitigate risks and capitalize on the dynamic nature of these markets.

It is important for investors to focus on identifying startups with strong potential in the highlighted LATAM countries in order to effectively utilize this ranking and framework. They can use the TAM, SAM, and SOM metrics alongside the DCF method and the ranking to make informed decisions about which startups to support. Additionally, collaborating with local governments and institutions to improve infrastructure and educational systems will enhance the investment climate. Investors should also consider establishing partnerships with local universities and tech hubs to foster innovation and talent development.

Future research should aim to enhance the valuation framework by including more detailed data on market conditions and startup performance across various countries in Latin America. Additionally, delving into specific opportunities within the tech industry, beyond the online food sector, could provide further insights into the diverse potential of AI applications. Conducting longitudinal studies to track the progress of AI-driven startups in the region and their impact on economic growth would also be valuable. Lastly, examining the role of policy and regulatory environments in shaping the success of tech startups in Latin America could offer crucial guidance for investors and policymakers.

%Bibliography
\bibliographystyle{unsrt}  
\bibliography{citations}

\end{document}